# A Unified Platform Enabling Power System Circuit Model Data Transfer Among Different Software

A. Khoshkbar-Sadigh, M. Heydari
M. Tedde, K. Smedley
Electrical Engineering and Computer Science Department
University of California-Irvine
CA, USA
a.khoshkbar.sadigh@ieee.org

Reza Arghandeh, Alexandra von Meier
California Institute for Energy and Environment (CIEE),
University of California-Berkeley
CA, USA

*Abstract*— **Diversity of software packages to simulate the power system circuits is considerable. It is challenging to transfer power system circuit model data (PSCMD) among different software tools and rebuild the same circuit in the second software environment. This paper proposes a unified platform (UP) where PSCMD are stored in a spreadsheet file with a defined format. Script-based PSCMD transfer applications, written in MATLAB, have been developed for a set of software to read the circuit model data from the UP spreadsheet and reconstruct the circuit in the destination software. This significantly eases the process of transferring circuit model data between each pair of software tools. In this paper ETAP, OpenDSS, Grid LabD, and DEW are considered. In order to test the developed PSCMD transfer applications, circuit model data of a test circuit and an actual sample circuit from a Californian utility company, both built in CYME, were exported into the spreadsheet file according to the UP format. Thereafter, circuit model data were imported successfully from the spreadsheet files into all above mentioned software using the PSCMD transfer applications developed for each software individually. Finally, load flow analysis is performed in all software and the obtained results match with each other.**

*Index Terms*— **power system simulation; power distribution; load flow analysis; power engineering; computer aided engineering.**

## I. INTRODUCTION

As power distribution systems are evolving into more complex networks, electrical engineers have to rely on software tools to perform circuit analysis [1]–[4]. Recent advances in engineering sciences have brought a revolution in power system software packages [5]–[11]. There are dozens of powerful software tools available in the market to simulate the power grid. Although their main functions are similar, there are differences in features and formatting structures to suit specific applications. This creates challenges for transferring circuit models between different software. Most utilities use some specific software package according to their needs or preferences, where each stores information about loads and circuit model in its own specific database structure.

With the emergence of new generation resources such as solar energy and wind energy, or new technologies such as power measurement units (PMUs) and micro PMUs [12]–[14], as well as new concepts such as smart grid or micro grid [15]–[18], new circuit phenomena including dynamic behaviors will need to be studied. Therefore, utilities may need to use different software tools to investigate these new phenomena, which may not be supported by their current software platforms. Thus, it becomes necessary to transfer power system circuit model data (PSCMD) from one software to another. However, PSCMD sharing among different software is a cumbersome process that can consume many person-hours. What is needed is a solution that enables cross-platform PSCMD transfer in the form of a complete power distribution network starting from the substation transformer all the way down to the load.

The objective of this paper is to develop a Unified Platform (UP) to facilitate transferring PSCMD among different software packages and relieve the challenges of the circuit model conversion process. UP uses a commonly available spreadsheet file with a defined format, for any home software to write data to, and for any destination software to read data from, via a script-based application called the PSCMD transfer application. The main considerations in developing the UP are to minimize manual intervention and import a one-line diagram into the destination software or export it from the source software, with all details to allow load flow, short circuit and other analyses.

In this paper, ETAP, OpenDSS, GridLab-D and DEW are considered. PSCMD transfer applications written in MATLAB have been developed for each of these to read the circuit model data provided in the UP spreadsheet. Each PSCMD transfer application has been verified by using two circuits, a test circuit and an actual circuit from a utility company for all the above listed software. When PSCMD is provided in the UP spreadsheet with defined format, successful reconstruction of the circuit in a destination software is achieved. Load flow analysis is performed in each software for both sample circuits and compared with the available results to verify the correctness of the circuit built by the PSCMD transfer application. The obtained results match accurately in all software for both circuits. The paper is organized as follows: Section II describes the UP and its functions. Section III presents simulation results to verify the effectiveness of the proposed UP and transfer process. Section IV compares features and capabilities of the mentioned software tools and highlights some specific issues, and Section V offers concluding observations.

## II. PROPOSED UNIFIED PLATFORM BASED ON SPREADSHEET

Conventionally, users develop their own tailor-made applications to transfer PSCMD from one software to another as needed. These applications are typically unidirectional, i.e. one specific application can transfer data from software #1 into software #2, but not vice versa. Therefore, if there are N different software packages,    applications will be required. This approach is quite wasteful since there are dozens of software packages in this field. To complicate things, each software has its own  terminology. For example, software #1 may name a positive sequence resistance of cable "Line R1 Ohms" while software #2 may name it "Pos. Seq R". Also each software has its own database structure to store their circuit model data. Another issue is the unit of component parameters. For example, conductor impedance can be in ohms or ohms per mile. Moreover, it is possible that one software package might consider certain parameters that other packages do not. Therefore, the developers of the PSCMD transfer applications need to be very familiar with both software and their internal languages (names of components and their parameter) in detail to be able to write the application properly. As noted earlier, different software may present This paper proposes UP to facilitate the PSCMD transfer process. As shown in Fig. 1, the proposed UP results in developing only two application for each software; one to read PSCMD from the UP spreadsheet file and transfer it to the destination software and the other one to read PSCMD from the source software and import it to the UP spreadsheet with defined format. Consequently, the number of applications required to share data among N different software is reduced from   to 2N. The PSCMD transfer applications that transform circuit model data from the UP spreadsheet to commercial software packages are written in the MATLAB. Each application is unique to the associated software.

The proposed UP consists of a spreadsheet file with a defined format containing several sheets to include specifications for bus/node, cable, capacitor bank, circuit breaker, generator, load, overhead line, transformer, etc. Each sheet is a library of the parameters of available components in the circuit with details to allow different power system studies. For instance, the sheet called cable contains the required parameters such as ID, from node, to node, phase configuration, positive sequence resistance, positive sequence reactance, positive sequence admittance, zero sequence resistance, zero sequence reactance, zero sequence admittance, length, current rating, etc. A snapshot of the UP spreadsheet is given in Fig. 2, illustrating the format and required parameters of some components such as transformer, cable, capacitor and load. It is worth mentioning that UP does not include the time-varying data to do time-series simulation since this depends on whether the specific software has the capability to do time-series simulation and if so, it depends on what format the specific software needs data to do time-series simulation. UP allows PSCMD transfer to and from the set of software, and it enables users to explore some new phenomena in their circuits by using a different software without the time consuming data transfer process. As mentioned previously, an individual application is developed in Matlab for each software tool including ETAP, OpenDSS, GridLAB-D and DEW to transfer PSCMD from UP into destination software.

It is worth mentioning that OpenDSS and GridLab-D are script-based tools where all components and their parameters and connectivity are assigned in the script-based environment. In other words, they do not have any Graphical User Interface (GUI) to drag and drop a component (e.g., load) from their library and make any connection to another component (e.g., cable). Moreover, they do not have any built-in feature yet to import PSCMD from a spreadsheet/access file, for example, as ETAP and DEW do. Therefore, it will take a long time to create any industry-level circuit in OpenDSS and GridLAB-D containing hundreds of lines, cables, loads etc. For instance, the general scheme of importing PSCMD from the UP spreadsheet into OpenDSS and GridLab-D using the developed applications is illustrated in Fig. 3.

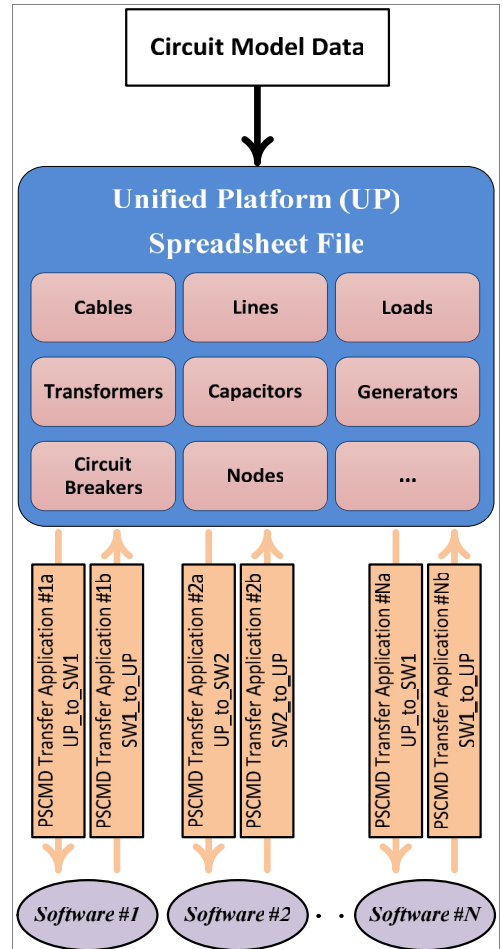

Fig. 1. Conceptual illustration of the proposed unified platform (UP) to transfer power system circuit model data among different software.  SW stands for software.

### (a) Transformer

| A | B | C | D | E | F | G | H | I | J | K | L |
|---|---|---|---|---|---|---|---|---|---|---|---|
| Section Id | From Node | To Node | Cap Nom (kVA) | Prim Volt (kVLL) | Sec Volt (kVLL) | X0 R0 Ratio | X1 R1 Ratio | Z0 (%) | Z1 (%) | Primary Config | Secondary Config |
| TR1 | N5 | N9 | 1500 | 12 | 0.48 | 10 | 10 | 6 | 6 | D | Yg |

### (b) Cable

| A | B | C | D | E | F | G | H | I | J | K | L | M | N |
|---|---|---|---|---|---|---|---|---|---|---|---|---|---|
| Section Id | From Node | To Node | Phase | Line R1 Ohms | Line X1 Ohms | Line B1 uS | Line R0 Ohms | Line X0 Ohms | Line B0 uS | Length ft | IA (Amps) | IB (Amps) | IC (Amps) |
| C1 | N1 | N2 | ABC | 0.01 | 0.012 | 12.091 | 0.054 | 0.015 | 12.091 | 341 | 473 | 473 | 473 |

### (c) Capacitor

| A | B | C | D | E | F | G | H | I | J | K |
|---|---|---|---|---|---|---|---|---|---|---|
| Section Id | From Node | Cap. Control | Cap. Status | kV | Total Cap. Kvar | Phase | Config | Sensing ON | Sensing OFF | PT Ratio |
| CAP2 | N4 | Voltage | On | 7.2 | 1200 | ABC | Y | 121 | 126 | 60 |

### (d) Load

| A | B | C | D | E | F | G | H | I | J |
|---|---|---|---|---|---|---|---|---|---|
| Section Id | From Node | Phase | Config | Spot kVAR A | Spot kVAR B | Spot kVAR C | Spot kW A | Spot kW B | Spot kW C |
| LD1 | N10 | AB | Yg | 600 | 360 | 0 | 800 | 480 | 0 |

Fig. 2. Snapshot of UP's spreadsheet containing required parameters of: (a) transformer; (b) cable; (c) capacitor and (d) load.

## III. SIMULATION RESULTS

In order to verify the effectiveness of the proposed UP and test the functionality of developed PSCMD transfer applications, the circuit model data of two different sample circuits were imported into ETAP, OpenDSS, GridLAB-D and DEW using the developed PSCMD transfer applications: (1) a test circuit built by authors as an example and (2) an actual sample circuit from a California utility. The load-flow analysis of structured circuits is performed in each software and the obtained results are compared with provided results from CYME to check the correctness of structured circuits. The following is general information about two imported circuits:

- The circuit model data of the test circuit contains 6 loads which are illustrated in Table I; two 1200-kVAR capacitor banks connected to nodes N4 and N8; 10 nodes (N1 to N10); 5 cables; one overhead line; and one transformer (12/0.48 kV) connected between nodes N5 and N9.
- The actual sample circuit from California utility contains 39 loads with total rating of 6.4 MW and 4 MVAR; four 1800-kVAR capacitor banks (two of which are connected to the grid in this case of study); 291 nodes; 119 cables; 36 overhead lines; 10 PV generators with total capacity of 5 MW; and 10 transformers (12/0.21 kV).

The obtained load flow results for both test and actual sample circuits are illustrated in Table II and Table III, respectively. We note that the obtained results match precisely with each other as well as results available from CYME, which confirms the correctness of the reconstructed circuit and the proper functionality of PSCMD transfer applications to transfer the circuit model data from UP spreadsheet to each software.

## IV. DISCUSSION

While the conversion process was successful for all four software tools in this paper, there are several differences among their features that merit discussion and are summarized in Table IV.

• ETAP and DEW have a GUI to drag and drop components from a library, while both OpenDSS and GridLAB-D have a script-based environment. Thus, users of OpenDSS and GridLAB-D need to write a text file according to the language of related software to define all components, parameters as well as their connectivity.

• ETAP has a useful feature to display a visual circuit schematic even if the XY coordinates of nodes are not imported into ETAP, which helps in the debugging stage. OpenDSS and DEW require XY coordinates to create a circuit schematic. GridLAB-D does not produce a schematic at all.

• GridLAB-D, OpenDSS, and DEW can model the line/cable based on ABC (phase-to-phase) format of line/cable impedances, while ETAP cannot. Moreover, ETAP, OpenDSS, and DEW can model the line/cable based on PNZ (positive-negative-zero) sequence format of line/cable impedances while GridLAB-D cannot.

• Software varies in its capability to model the capacitance of lines/cables, which can be important in distribution level circuit studies since the capacitance of underground cable is considerable and may have a significant impact on the load flow results. ETAP, DEW and OpenDSS have this capability while GridLAB-D does not. According to the simulation results shown in Table II and III, however, GridLAB-D load flow results nevertheless match with the others.. The reason for this is that the capacitances of lines/cables are imported as individual capacitors (such as load) at the beginning and ending-point of lines/cables in the PSCMD transfer application.

• The voltage rating of all nodes should be assigned one by one in GridLAB-D, while ETAP, OpenDSS and DEW get it automatically after connecting the main feeding point and assigning the connectivity of the circuit.

TABLE I: LOAD SUMMARY OF TEST CIRCUIT.

| Name | Node | kW | kVAR | Voltage Rating | Power Factor (%) |
|---|---|---|---|---|---|
| LD1 | N4 | 3000 | 1800 | 7.2 kV | 80 |
| LD2 | N5 | 4500 | 2700 | 7.2 kV | 80 |
| LD3 | N7 | 3000 | 1307.7 | 7.2 kV | 90 |
| LD4 | N8 | 1500 | 653.7 | 7.2 kV | 90 |
| LD5 | N9 | 1500 | 900 | 0.48 kV | 80 |
| LD6 | N10 | 3000 | 1800 | 7.2 kV | 80 |

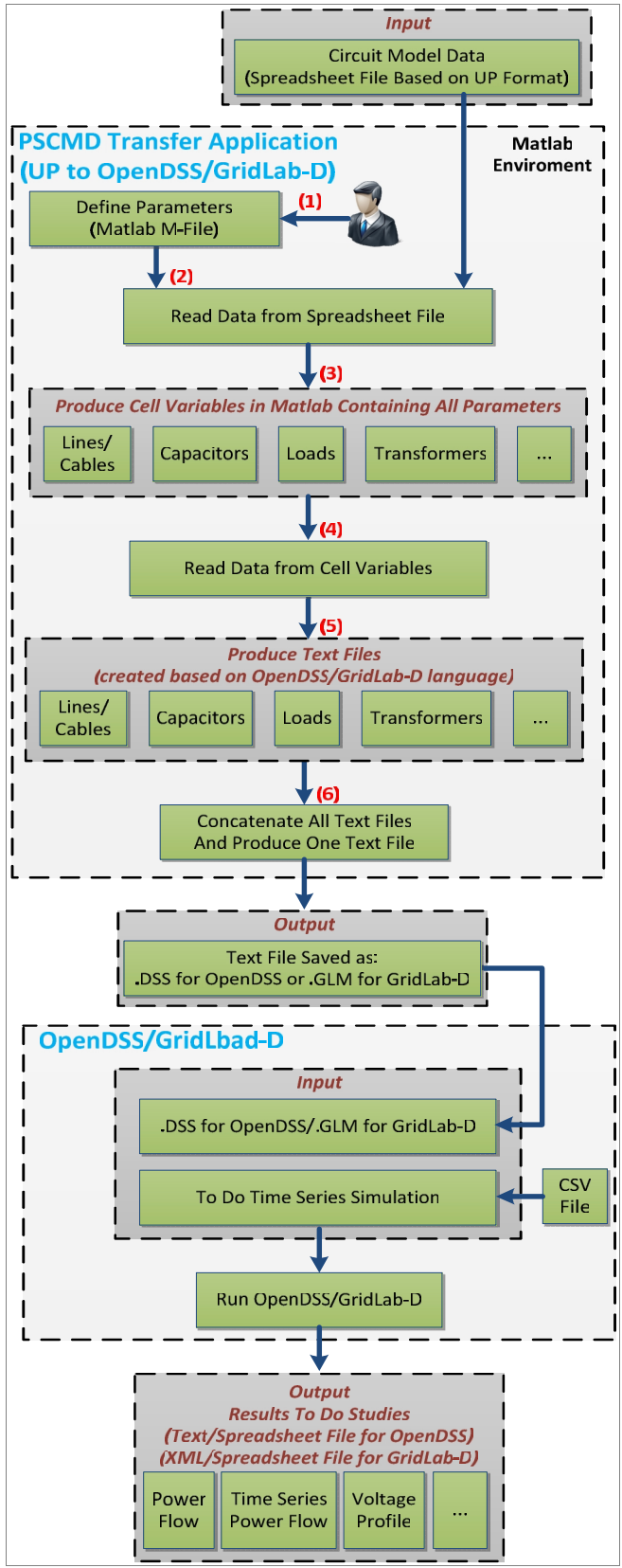

Fig. 7. General scheme of importing power system circuit model data from UP spreadsheet into OpenDSS/GridLab-D using PSCMD transfer application.

## V. CONCLUSION

Emerging technologies are introducing new devices and phenomena to power distribution circuits, creating a need for new software capabilities to investigate them since the existing software may not support them. Consequently, users may need to transfer their PSCMD among different software tools to take advantage of specific features in new software. Due to the considerable diversity of software tools on the market to simulate and study power system circuits, it has been a challenge to transfer PSCMD and reconstruct circuits in different software tools. With the UP scheme proposed in this paper, all PSCMD are presented in a spreadsheet file based on a single defined format. As a result, only two unidirectional applications are required to transfer PSCMD from the UP spreadsheet into any one software and vice versa, instead of a combinatorial number of conversion applications between multiple software. In this paper, PSCMD transfer applications were developed in MATLAB for ETAP, OpenDSS, GridLAB-D and DEW and tested on two circuits to confirm that the load flow results agree and the circuit conversions have been successful.

The conversion applications presented in this paper is available to utility engineers and researchers to facilitate their work. Future development may expand the conversion applications to include other commonly used software tools.

ACKNOWLEDGEMENT

This work was supported by the California Energy Commission under CEC Contract #500-11-019, also known as "Distribution Monitoring for Renewables Integration (DMRI)." The authors would like to thank all the participating utilities in the DMRI project, especially, Sunil Shah from Southern California Edison (SCE) and Steve Kirin from Sacramento Municipal Utility District (SMUD) for their time and effort to provide circuit model data. Appreciation also goes to Ken Nunes from the San Diego Super Computer Center (SDSC) to facilitate data transfer, James Sherwood from Rocky Mountain Institute for his valuable and constructive suggestions and ETAP Inc. for providing an ETAP software license.

REFERENCES

[1] V. Dinavahi and Y. Chen, "Multi-FPGA digital hardware design for detailed large-scale real-time electromagnetic transient simulation of power systems," *IET Gener. Transm. Distrib.*, vol. 7, no. 5, pp. 451–463, May 2013.

[2] D. Fabozzi, A. S. Chieh, B. Haut, and T. Van Cutsem, "Accelerated and Localized Newton Schemes for Faster Dynamic Simulation of Large Power Systems," *IEEE Trans. Power Syst.*, vol. 28, no. 4, pp. 4936–4947, Nov. 2013.

[3] M. Fan, V. Ajjarapu, C. Wang, D. Wang, and C. Luo, "RPM-Based Approach to Extract Power System Steady State and Small Signal Stability Information From the Time-Domain Simulation," *IEEE Trans. Power Syst.*, vol. 26, no. 1, pp. 261–269, Feb. 2011.

[4] V. Jalili-Marandi, Z. Zhou, and V. Dinavahi, "Large-Scale Transient Stability Simulation of Electrical Power Systems on Parallel GPUs," *IEEE Trans. Parallel Distrib. Syst.*, vol. 23, no. 7, pp. 1255–1266, Jul. 2012.

[5] A. Yazdani, A. R. Di Fazio, H. Ghoddami, M. Russo, M. Kazerani, J. Jatskevich, K. Strunz, S. Leva, and J. A. Martinez, "Modeling Guidelines and a Benchmark for Power System Simulation Studies

of Three-Phase Single-Stage Photovoltaic Systems," *IEEE Trans. Power Deliv.*, vol. 26, no. 2, pp. 1247–1264, Apr. 2011.

[6] R. D. Zimmerman, C. E. Murillo-Sanchez, and R. J. Thomas, "MATPOWER: Steady-State Operations, Planning, and Analysis Tools for Power Systems Research and Education," *IEEE Trans. Power Syst.*, vol. 26, no. 1, pp. 12–19, Feb. 2011.

[7] Y. Zou, M. E. Elbuluk, and Y. Sozer, "Simulation Comparisons and Implementation of Induction Generator Wind Power Systems," *IEEE Trans. Ind. Appl.*, vol. 49, no. 3, pp. 1119–1128, May 2013.

[8] S. Section, V. Large, and P. Systems, "Foreword for the Special Section on Analysis and Simulation of Very Large Power Systems," *IEEE Trans. Power Syst.*, vol. 28, no. 4, pp. 4885–4887, Nov. 2013.

[9] K. Anderson, J. Du, A. Narayan, and A. El Gamal, "GridSpice: A distributed simulation platform for the smart grid," in *2013 Workshop on Modeling and Simulation of Cyber-Physical Energy Systems (MSCPES)*, 2013, pp. 1–5.

[10] I. Nagel, L. Fabre, M. Pastre, F. Krummenacher, R. Cherkaoui, and M. Kayal, "High-Speed Power System Transient Stability Simulation Using Highly Dedicated Hardware," *IEEE Trans. Power Syst.*, vol. 28, no. 4, pp. 4218–4227, Nov. 2013.

[11] D. Wang, B. de Wit, S. Parkinson, J. Fuller, D. Chassin, C. Crawford, and N. Djilali, "A test bed for self-regulating distribution systems: Modeling integrated renewable energy and demand response in the GridLAB-D/MATLAB environment," in *2012 IEEE PES Innovative Smart Grid Technologies (ISGT)*, 2012, pp. 1–7.

[12] M. Dehghani, L. Goel, and W. Li, "PMU based observability reliability evaluation in electric power systems," *Electr. Power Syst. Res.*, vol. 116, pp. 347–354, Nov. 2014.

[13] J. E. Anderson and A. Chakrabortty, "PMU placement for dynamic equivalencing of power systems under flow observability constraints," *Electr. Power Syst. Res.*, vol. 106, pp. 51–61, Jan. 2014.

[14] A. Mahari and H. Seyedi, "Optimal PMU placement for power system observability using BICA, considering measurement redundancy," *Electr. Power Syst. Res.*, vol. 103, pp. 78–85, Oct. 2013.

[15] A. Mohamed and O. Mohammed, "Real-time energy management scheme for hybrid renewable energy systems in smart grid applications," *Electr. Power Syst. Res.*, vol. 96, pp. 133–143, Mar. 2013.

[16] S. Goleijani, T. Ghanbarzadeh, F. Sadeghi Nikoo, and M. Parsa Moghaddam, "Reliability constrained unit commitment in smart grid environment," *Electr. Power Syst. Res.*, vol. 97, pp. 100–108, Apr. 2013.

[17] C. Battistelli and A. J. Conejo, "Optimal management of the automatic generation control service in smart user grids including electric vehicles and distributed resources," *Electr. Power Syst. Res.*, vol. 111, pp. 22–31, Jun. 2014.

[18] S. Kahrobaee, R. A. Rajabzadeh, L.-K. Soh, and S. Asgarpoor, "Multiagent study of smart grid customers with neighborhood electricity trading," *Electr. Power Syst. Res.*, vol. 111, pp. 123–132, Jun. 2014.

TABLE II: LOAD FLOW RESULTS OF TEST CIRCUIT.

| Bus | Active Power (kW) | | | | | Reactive Power (kVAR) | | | | |
|---|---|---|---|---|---|---|---|---|---|---|
| | Original (CYME) | ETAP | OpenDSS | GridLAB-D | DEW | Original (CYME) | ETAP | OpenDSS | GridLAB-D | DEW |
| N1 | 14527 | 14536 | 14534 | 14535 | 14528 | 7982 | 7989 | 7988.7 | 7988.6 | 7986 |
| N2 | 14510 | 14518 | 14516.8 | 14517 | 14513 | 7963 | 7969 | 7969.4 | 7969.2 | 7968 |
| N3 | 14510 | 14518 | 14516.8 | 14517 | 14517 | 7963 | 7969 | 7969.4 | 7969.2 | 7969 |
| N4 | 14501 | 14509 | 14508 | 14508.6 | 14508 | 7954 | 7960 | 7961.4 | 7961.25 | 7957 |
| N5 | 8925 | 8931 | 8930.4 | 8931.51 | 8927 | 4852 | 4857 | 4857.2 | 4857.65 | 4855 |
| N6 | 4117 | 4120 | 4119.6 | 4120.16 | 4115 | 1141 | 1144 | 1144.1 | 1144.17 | 1144 |
| N7 | 4050 | 4053 | 4052.4 | 4052.9 | 4053 | 1045 | 1048 | 1047.7 | 1047.94 | 1047 |
| N8 | 1349 | 1350 | 1349.8 | 1350 | 1349 | -265 | -263 | -263.1 | -263.5 | -263 |
| N9 | 1199 | 1200 | 1199.9 | 1200 | 1200 | 899 | 900 | 899.9 | 900 | 900 |
| N10 | 2400 | 2400 | 2400 | 2400 | 2400 | 1800 | 1800 | 1800 | 1800 | 1800 |

TABLE III: PARTIAL LOAD FLOW RESULTS FROM ACTUAL SAMPLE CIRCUIT.

| Bus | Active Power (kW) | | | | | Reactive Power (kVAR) | | | | |
|---|---|---|---|---|---|---|---|---|---|---|
| | Original (CYME) | ETAP | OpenDSS | GridLAB-D | DEW | Original (CYME) | ETAP | OpenDSS | GridLAB-D | DEW |
| 90091686_02658 (c104) | 276 | 277 | 276.7 | 276.7 | 272 | 167 | 167 | 166.3 | 166.8 | 166 |
| PME4896-3_02658 (sw123) | 357 | 357 | 356.9 | 356.9 | 358 | 224 | 223 | 223.1 | 223.24 | 222 |
| 833E_02658 (sw94) | 380 | 380 | 380 | 380.1 | 378 | 238 | 237 | 237 | 237.2 | 239 |
| 107988591_02658 (c95) | 403 | 402 | 402 | 402.3 | 401 | 252 | 253 | 252.8 | 252.9 | 254 |
| GS0713-2_02658 (sw46) | 449 | 447 | 447 | 447 | 448 | 273 | 275 | 264.9 | 271.25 | 263 |
| PME4896-4_02658 (C73) | 550 | 552 | 550 | 549.2 | 554 | -1348 | -1349 | -1358 | -1363 | -1356 |
| J057-1P_02658 (SW113) | 576 | 576 | 575.5 | 575.5 | 575 | 360 | 360 | 359.3 | 359.3 | 361 |
| PME5100-1_02658 (C25) | 805 | 806 | 805.6 | 805.6 | 811 | -1233 | -1226 | -1232.5 | -1225.4 | -1229 |
| RCSG777-3_02658 (sw111) | 833 | 827 | 832.1 | 832.6 | 836 | 1098 | 1101 | 1103.9 | 1105.1 | 1099 |
| PMH5099-3_02658 (sw60) | 1106 | 1105 | 1105.4 | 1105.5 | 1104 | 693 | 693 | 692.8 | 693.18 | 689 |
| 48201834_02658 (c92) | 1255 | 1253 | 1252.8 | 1252.8 | 1253 | -970 | -960 | -967.6 | -958.8 | -966 |
| PMH5099-4_02658 (c20) | 1717 | 1716 | 1715.5 | 1715.7 | 1719 | -2419 | -2402 | -2415.9 | -2400 | -2398 |
| RCS5187-4_02658 (c19) | 2827 | 2826 | 2825.7 | 2826.3 | 2825 | -1730 | -1712 | -1727.1 | -1710 | -1707 |
| RCS5187-3_02658 (sw16) | 3150 | 3147 | 3146.7 | 3147.3 | 3153 | -1528 | -1509 | -1524.4 | -1507.4 | -1314 |
| 148639376_02658 (c89) | 3584 | 3584 | 3581.2 | 3583.7 | 3748 | -1301 | -1278 | -1299.2 | -1274.2 | -1294 |
| PS0372_02658 (sw7) | 4207 | 4209 | 4205.5 | 4208.7 | 4211 | -912 | -893 | -914.9 | -888 | -904 |
| 02658 | 3723 | 3726 | 3720.9 | 3729 | 3725 | 542 | 557 | 547.8 | 585 | 574 |